\documentclass[conference,10pt]{IEEEtran}

\ifCLASSINFOpdf
\else
\fi
\usepackage{tikz}
\usepackage{pgfplots}
\usepgfplotslibrary{units}

\usepackage{array}
\ifCLASSOPTIONcompsoc
  \usepackage[caption=false,font=normalsize,labelfont=sf,textfont=sf]{subfig}
\else
  \usepackage[caption=false,font=footnotesize]{subfig}
\fi

\usepackage{booktabs}
\usepackage{color}

\definecolor{Set1-7-1}{RGB}{228,26,28}
\definecolor{Set1-7-2}{RGB}{55,126,184}
\definecolor{Set1-7-3}{RGB}{77,175,74}
\definecolor{Set1-7-4}{RGB}{152,78,163}
\definecolor{Set1-7-5}{RGB}{255,127,0}
\definecolor{Set1-7-6}{RGB}{166,86,40}
\definecolor{Set1-7-7}{RGB}{0,0,0}
\definecolor{babyblue}{rgb}{0.54, 0.81, 0.94}

\newcommand{\figurewidth}{0.97}	% Width of plots as a fraction of the column width
\newcommand{\figureheight}{0.73} % Height of plots as a fraction of the column width

\usepackage[cmex10]{amsmath}
\usepackage{amssymb,amsthm}
\usepackage{mathrsfs}
\usepackage{amsbsy}
\usepackage{steinmetz}
\usepackage{nicefrac}
\usepackage{soul}
\usepackage{multirow}
\usepackage{graphicx}
\usepackage{epstopdf}
\usepackage{amsfonts}
\usepackage{xcolor}
\makeatletter
\def\testclr#1#{\@testclr{#1}}
\def\@testclr#1#2{{\fboxsep\z@\fbox{\colorbox#1{#2}{\phantom{XX}}}}}

\makeatother

\begin{document}

\title{An Open-Source LoRa Physical Layer Prototype on GNU Radio}

\author{\IEEEauthorblockN{Joachim Tapparel,\IEEEauthorrefmark{1} Orion Afisiadis,\IEEEauthorrefmark{1} Paul Mayoraz,\IEEEauthorrefmark{1} Alexios Balatsoukas-Stimming,\IEEEauthorrefmark{2}and Andreas Burg\IEEEauthorrefmark{1}}
\IEEEauthorblockA{\IEEEauthorrefmark{1}Telecommunication Circuits Laboratory, \'{E}cole polytechnique f\'{e}d\'{e}rale de Lausanne, Switzerland\\
\IEEEauthorrefmark{2}Department of Electrical Engineering, Eindhoven University of Technology, The Netherlands}%
}

\maketitle

\begin{abstract}
LoRa is the proprietary physical layer (PHY) of LoRaWAN, which is a popular Internet-of-Things (IoT) protocol enabling low-power devices to communicate over long ranges. A number of reverse  engineering  attempts  have  been  published  in  the  last  few  years that helped to reveal many of the LoRa PHY details.  In this work, we describe our standard compatible LoRa PHY software-defined radio (SDR) prototype based on GNU Radio. We show how this SDR prototype can be used to develop and evaluate receiver algorithms for LoRa. As an example, we describe the sampling time offset and the carrier frequency offset estimation and compensation blocks. We experimentally evaluate the error rate of LoRa, both for the uncoded and the coded cases, to illustrate that our publicly available open-source implementation is a solid basis for further research.
\end{abstract}

\begin{IEEEkeywords}
LoRa, GNU Radio, Internet-of-Things
\end{IEEEkeywords}

\IEEEpeerreviewmaketitle

\section{Introduction}

LoRa is one of the most popular low-power wireless standards for the Internet of Things (IoT)~\cite{Boulogeorgos2016} and it has attracted significant attention both in the industry and in academia~\cite{Haxhibeqiri2018,Vangelista2017}. The specification of the LoRa physical layer (PHY) is proprietary, making its study challenging. Nevertheless, performance and robustness studies have been performed using commercial LoRa transceivers~\cite{Patel2017,Lone2018,Callebaut2019,Fernandes2019}. The disadvantage of this approach is that it does not allow to evaluate modifications of the LoRa PHY or more advanced receiver algorithms, since these are pre-defined and fixed in commercial transceivers.

Software-defined radios (SDRs) can readily address this issue~\cite{Sklivanitis2016}, provided that a complete implementation of the LoRa PHY is available. Numerous reverse engineering efforts and corresponding SDR implementations (e.g.,~\cite{Knight2016,Robyns2018,MyriadRF,Marquet2019}) from the past few years have revealed many important details of the LoRa PHY. However, despite their  important contributions, these implementations focus on the reverse engineering aspect and are therefore using only basic receivers, which lack, for example, sampling time offset (STO) and carrier frequency offset (CFO) estimation and correction. This can have a devastating effect on the demodulation of LoRa symbols. As such, the existing SDR implementations can only operate at very high signal-to-noise ratios (SNRs).

\subsubsection*{Contributions}
In this work, we describe a fully-functional GNU Radio SDR implementation of a LoRa transceiver with all the necessary receiver components to operate correctly even at very low SNRs. We use variants of the algorithms described in~\cite{Bernier2019,Xhonneux2019} to jointly correct the STO and CFO. Moreover, our implementation fixes several minor issues of existing implementations (e.g., the order of channel coding and whitening), which are explained in more detail in~\cite{Tapparel2019} and are not explained again here due to space limitations. Finally, we provide, to the best of our knowledge, the first fully end-to-end experimental performance results of a LoRa SDR receiver at low SNR. Our GNU Radio implementation of the complete LoRa transceiver chain is publicly available at~\cite{TCLoRaRepo}.

\section{LoRa PHY}\label{sec:loraphy}
In this section, we first describe the LoRa modulation and the complete transmitter and receiver chains, including whitening, channel coding, interleaving, and Gray mapping and we then describe the structure of LoRa packets.

\subsection{LoRa Modulation and Demodulation}
LoRa is a spread-spectrum frequency modulation with bandwidth $B \in \{125,250,500\}$~kHz. Every LoRa symbol carries $\text{SF} \in \{7, \dots, 12\}$ bits and consists of $N = 2^\text{SF}$ chips, where $\text{SF}$ is called the \emph{spreading factor}. A baseband symbol $s \in \mathcal{S}\triangleq\left\{0,\hdots,N{-}1\right\}$ spans the entire bandwidth, i.e., the symbol begins at frequency $(\frac{s B}{N} - \frac{B}{2})$ and its frequency increases by $\frac{B}{N}$ in every chip. When the frequency $\frac{B}{2}$ is reached, a frequency fold to $-\frac{B}{2}$ occurs and the frequency continues to increase by $\frac{B}{N}$ in every following chip of the symbol until the initial frequency is reached. 
The discrete-time baseband-equivalent description of a LoRa symbol $s$, can be written in two forms. The first form does not ensure inter-symbol phase continuity since the initial phase of every symbol depends on the symbol itself. In the second form, which is used in~\cite{Ghanaatian2019,Afisiadis2019b,Elshabrawy2019b,Chiani2019}, the phase remains continuous between consecutive symbols. This phase continuity is highlighted as a useful property in one of the original LoRa patents~\cite{Seller2016}. Using the form that ensures inter-symbol phase continuity and for the case where the sampling frequency $f_s$ is equal to $B$, the baseband-equivalent representation of a LoRa symbol is~\cite{Afisiadis2019b}
\begin{align} \label{eq:LoRa_symbol}
	x_{s}[n] & = e^{j2\pi \left(\frac{n^2}{2N}  + \left(\frac{s}{N} - \frac{1}{2}\right)n \right)}, \;\; n \in \mathcal{S}.
\end{align}
Note that, as explained in detail in~\cite{Chiani2019}, the continuous-time LoRa chirp occupies a bandwidth that is slightly larger than B. Therefore, using a sampling rate $f_s = B$ in a real transmission introduces some distortion effects due to aliasing. %In order to avoid the distortion effects a higher bandwidth than B should be used ensured before sampling~\cite{Chiani2019}.
\begin{figure}[t]
	\centering
	\includegraphics[width=0.48\textwidth]{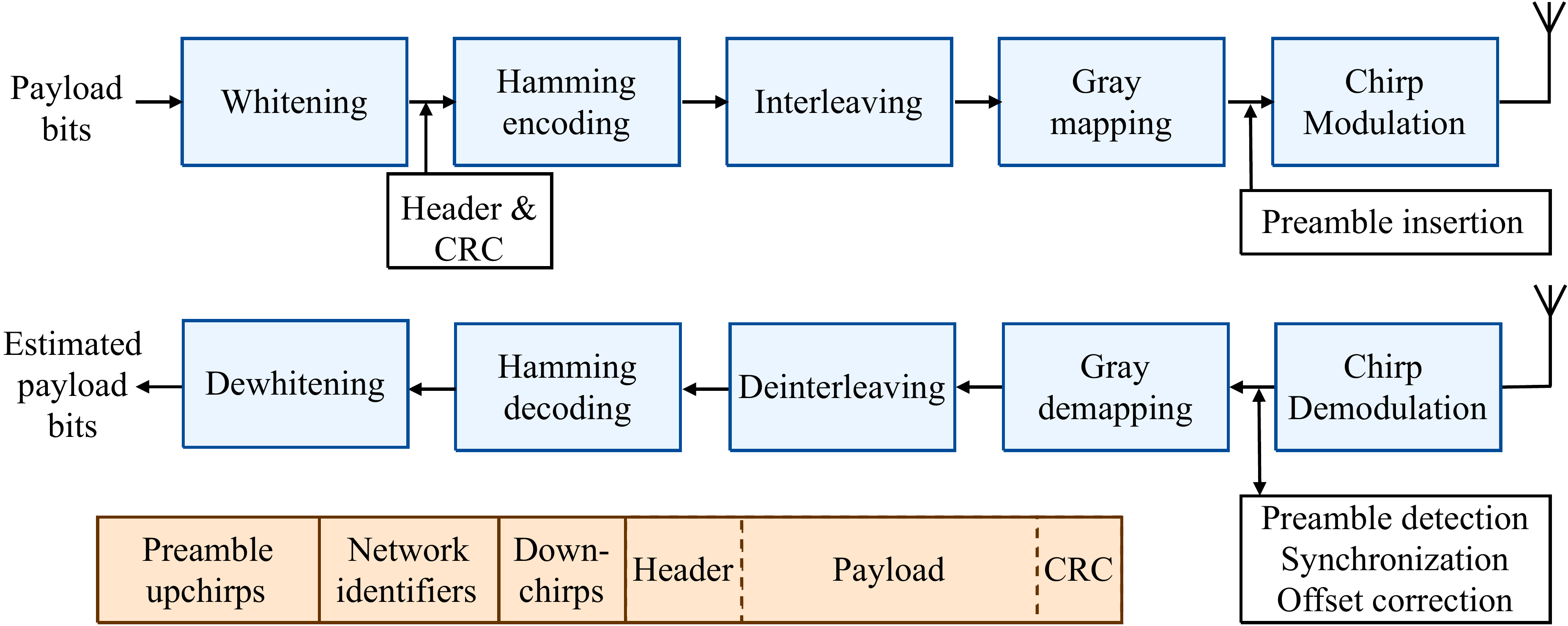}
	\caption{The LoRa PHY Tx and Rx chains and of the LoRa packet structure.}
	\label{fig:blockDiagram}
	\vspace{-0.25cm}
\end{figure}

When transmission takes place over an AWGN channel, the received LoRa signal is given by
\begin{align}
  y[n] & = x_{s}[n] + z[n], \;\; n \in \mathcal{S}, \label{eq:lora_rx}
\end{align}
where $z[n] \sim \mathcal{CN}(0,\sigma^2)$ is complex-valued AWGN with variance $\sigma^{2} = \frac{N_0}{2N}$ and singled-sided noise power spectral density $N_0$. In this case, the SNR is defined as $\text{SNR} = \frac{1}{N_0}$.

The receiver first \emph{dechirps} the received signal by multiplying $y[n]$ with the complex conjugate of a reference signal $x_{0}[n]$, which is typically chosen to be the LoRa symbol for $s=0$ (i.e., a pure \emph{upchirp}) for simplicity. After that, the receiver computes a discrete Fourier transform (DFT), obtaining $\mathbf{Y} = \text{DFT}\left(\mathbf{y} \odot \mathbf{x}_{0}^{*}\right)$, where $\mathbf{y} = [ y[0] \hdots y[N{-}1]]$ and $\mathbf{x}_{0} = [ x_{0}[0] \hdots x_{0}[N{-}1]]$ and where $\odot$ denotes element-wise vector multiplication. In a typical non-coherent LoRa receiver, the estimate $\hat{s}$ of the transmitted symbol is
\begin{align}
  \hat{s} = \arg\max_{k\in \mathcal{S} } \left( |Y[k]| \right). \label{eq:retrieved_symbol_dft}
\end{align}
Each demodulated LoRa symbol $\hat{s}$ has an associated $\text{SF}$-bit label, which gives an estimate of the $\text{SF}$ transmitted bits.

\subsection{Complete LoRa Transceiver Chain}
Apart from the modulation and demodulation explained in the previous section, the transmitter and receiver chains also perform some additional processing, as shown in Fig.~\ref{fig:blockDiagram}. The LoRa transmitter chain performs whitening, Hamming encoding, interleavering, and Gray mapping prior to the chirp modulation. The receiver performs Gray demapping, deinterleaving, Hamming decoding, and dewhitening.

\subsubsection{Whitening}
Whitening is an XOR of the information bits with a pseudo-random sequence. The whitening sequence that LoRa uses cna be found in~\cite{TCLoRaRepo} and was derived from the whitening matrix shown in~\cite[Fig.~2.11]{Tapparel2019}. The base whitening matrix we found is the one that corresponds to the coding rate (CR) $\text{CR} = \nicefrac{4}{8}$. For $\text{CR}=\nicefrac{4}{7}$ and $\text{CR}=\nicefrac{4}{6}$ one and two of the rightmost columns of the base whitening matrix are removed, respectively. For $\text{CR} = \nicefrac{4}{5}$ the last column in the whitening matrix is different than the corresponding fifth column in the base whitening matrix. This happens because, as explained in the sequel, for $\text{CR} = \nicefrac{4}{5}$ a single parity check is used instead of a Hamming code. Moreover, we found that for the coding rate $\text{CR} = \nicefrac{4}{5}$, the parity bit is calculated from the whitened version of the four bits, meaning that the whitening block is the first block in the transmitter chain. 

\subsubsection{Error-Correction Coding}
LoRa supports four error-correction coding (ECC) rates $\text{CR} \in \left\{\nicefrac{4}{5},\nicefrac{4}{6},\nicefrac{4}{7},\nicefrac{4}{8}\right\}$. For the three lowest rates, LoRa uses $(k,n)$ Hamming codes with $ k = 4 $ and $n \in \{6,7,8\}$, where $k$ is the data length and $n$ is the codeword length. The $(4,6)$ Hamming code is a punctured version of the standard $(4,7)$ Hamming code, while the $(4,8)$ Hamming code is an extended version of the $(4,7)$ Hamming code. For $\text{CR} = \nicefrac{4}{5}$, LoRa uses an even parity-check code instead of a (punctured) $(4,5)$ Hamming code. The generator and parity-check matrices, are described in detail in~\cite{Tapparel2019}. 

\subsubsection{Interleaving}
LoRa uses a diagonal interleaver to distribute the (up to $\text{SF}$) bit errors resulting from a symbol error over multiple ECC codewords. The combination of interleaving with the ECC leads to a higher probability of correctly decoded codewords since most codewords will only contain a single bit error, as explained in detail in~\cite{Afisiadis2019c}.

\subsubsection{Gray mapping}
LoRa uses a reverse Gray code for the mapping from bits to symbols. Thus, a symbol error that mistakes a symbol for one of its adjacent symbols (also called a \emph{$\pm 1$ demodulation error}) only causes \emph{a single} bit error, which can always be corrected by the Hamming codes with $\text{CR} = \nicefrac{4}{7}$ and $\text{CR} = \nicefrac{4}{8}$. This property is particularly useful if CFO or STO can not fully be recovered, which typically leads to $\pm 1$ demodulation errors~\cite{Seller2016,Bernier2019,Xhonneux2019,Afisiadis2019c}.

\subsection{LoRa Packet Structure}
The structure of a LoRa packet, which has been explained in detail in~\cite{Robyns2018}, is shown in Fig.~\ref{fig:blockDiagram}. 

\subsubsection{Preamble}
The first part of a LoRa packet is the preamble, which consists of a variable number $ N_{\text{pr}} $ of upchirps. The default value for $ N_{\text{pr}} $ is 8, but all preamble lengths in the range $ N_{\text{pr}} \in \{6,\hdots,65535\}$ are valid~\cite{SX127x}, enabling effective preamble detection for a very large range of effective SNRs. 
\subsubsection{Network identifiers}
After the preamble, the packet contains two network identifier symbols. In~\cite{Seller2016} it is mentioned that the network identifier symbols are modulated as $\{x,N-x\}$, where $x$ is the network identifier, and that they should have a minimum distance of three for different networks to avoid problems caused by $\pm 1$ demodulation errors. However, it is also mentioned in~\cite{Bernier2019} that the network identifiers observed were actually of the form $\{x,x\}$. 
\subsubsection{Downchirps}
After the network identifiers, there are two and a quarter frequency synchronization symbols, which are downchirps $ \mathbf{x}_{0} ^{*}$. These downchirps are used in~\cite{Bernier2019,Xhonneux2019} to partially distinguish between the STO and the CFO.
\subsubsection{Header (Optional)}
The packet continues with an optional header, which contains information about the length of the packet, the code rate, the presence of a cyclic redundancy check (CRC), and a checksum. If the header is not present, the receiver parameters need to be configured manually. The structure of the header is explained in detail in~\cite{Robyns2018} and~\cite{Tapparel2019}.
\subsubsection{Payload and CRC}
Finally, the last part of the packet is the payload and an optional 16-bit CRC of the payload bits~\cite{SX127x}. The maximum length of the payload is 255 bytes.

\section{LoRa Frame Synchronization} \label{sec:frameSynch}
The demodulation procedure described in Section~\ref{sec:loraphy} assumes that the receiver is perfectly synchronized to the incoming signal and that no impairments are present. However, in practice these assumptions typically do not hold and some additional processing steps are required in order to ensure correct frame synchronization and to correct the STO and CFO impairments. We note that, in LoRa, the STO and CFO need to be corrected already during the frame synchronization step and not only for the data part of the frame. In this section, we describe how we ensure fine-grained synchronization and correct demodulation in the presence of STO and CFO in our USRP-based GNU Radio LoRa receiver. To this end, we first write an expression for the received signal before frame synchronization, in order to later describe the appropriate frame synchronization and offset correction algorithms.

\subsection{Frame Synchronization Signal Model}
The received packet is sampled with a frequency $f_{s}$ and decisions are taken in windows of $N$ samples. The decision windows are generally not synchronized to the preamble upchirps and the $N$ samples of a decision window will typically consist of parts of two preamble upchirps, which we denote by $s_{\text{up}_{1}}$ and $s_{\text{up}_{2}}$ and where $s_{\text{up}_{1}} = s_{\text{up}_{2}} = 0$. Following a similar notation as~\cite{Afisiadis2019b}, let $\tau_{\text{STO}}$ be the relative time offset between the first chip of the decision window and the first chip of $s_{\text{up}_{2}}$ in the decision window. %Since there is no coordination between the transmitter and receiver, $\tau_{\text{STO}}$ generally has a uniform distribution in $[0, N)$. 
Then the discrete-time baseband-equivalent equation of the transmitted signal $x[n]$ at the receiver is~\cite[Eq. (21)]{Afisiadis2019b}
\begin{align}
x[n] & =
\begin{cases}
e^{j2\pi \left(\frac{(n + N-\tau_{\text{STO}})^{2}}{2N} -\frac{1}{2}(n + N-\tau_{\text{STO}})\right)}, & n \in \mathcal{N}_{L_1},\\
e^{j2\pi \left(\frac{(n - \tau_{\text{STO}})^{2}}{2N} -\frac{1}{2}(n - \tau_{\text{STO}})\right)}, & n \in \mathcal{N}_{L_2},
\end{cases}
\end{align}
where $ \mathcal{N}_{L_1} = \{0, \dots,\lceil \tau_{\text{STO}}\rceil-1\}$ and $ \mathcal{N}_{L_2} = \{\lceil \tau_{\text{STO}}\rceil, \dots, N-1 \} $.
Moreover, let $ f_{c_1} $ denote the carrier frequency used during up-conversion at the transmitter and $ f_{c_2} $ the carrier frequency used during down-conversion at the receiver. The carrier frequency offset is the difference \mbox{$\Delta f_{c} = f_{c_1} {-} f_{c_2}$}. The corresponding signal model is
\begin{align}
y[n] & = hc[n]x[n] + z[n], \;\; n \in \mathcal{S}, \label{eq:lora_rx_int}
\end{align}
where $h$ is the channel gain between the transmitter and the receiver, $ c[n] = e^{j2\pi (n{+}(m-1)N) \frac{\Delta f_{c}}{f_{s}}}$ is the CFO term affecting the $m$-th symbol in the packet, $x[n]$ is the transmitted signal, and $z[n] \sim \mathcal{N}(0,\sigma^2)$ is AWGN. The STO $\tau_{\text{STO}}$ can be separated into an integer part $L_{\text{STO}} = \left \lfloor{\tau_{\text{STO}}}\right \rfloor$, and a fractional part $\lambda_{\text{STO}} = \tau_{\text{STO}} - \left \lfloor{\tau_{\text{STO}}}\right \rfloor$. In addition, the CFO translates into an offset $\tau_{\text{CFO}} = \frac{\Delta f_{c}N}{f_{s}}$~\cite{Afisiadis2019c}, which can also be split into an integer part $L_{\text{CFO}} = \left \lfloor{\tau_{\text{CFO}}}\right \rfloor$, and a fractional part $\lambda_{\text{CFO}} = \tau_{\text{CFO}} - \left \lfloor{\tau_{\text{CFO}}}\right \rfloor$. The STO and the CFO affect the signal in a combined manner that makes their joint estimation and correction challenging~\cite{Bernier2019,Xhonneux2019}.

\subsection{Synchronization and Offset Correction}

\subsubsection{Preamble detection}\label{sec:preambledet}

In the presence of STO and CFO and in the absence of AWGN, it can be shown that the demodulation decisions of the detection windows during the preamble upchirps are $\hat{s} = \lfloor \tau_{\text{STO}} - \tau_{\text{CFO}} \rceil = \lfloor \left(L_{\text{STO}} - L_{\text{CFO}} \right)  + \left(\lambda_{\text{STO}}-\lambda_{\text{CFO}}\right)  \rceil$. Our SDR LoRa receiver detects the presence of a preamble when $N_{\text{pr}}{-}1$ consecutive symbols are demodulated with values in a range $\{s{-}1,s,s{+}1\}$. The reason for allowing this margin during preamble detection is that  the fractional offsets $\lambda_{\text{STO}}{-}\lambda_{\text{CFO}}$ can lead to $\pm 1$ demodulation errors in the presence of AWGN~\cite{Seller2016,Ghanaatian2019,Bernier2019,Xhonneux2019}. Finally, the preamble synchronization value $\hat{s}_{\text{pr}}$ is decided using a majority rule from the $N_{\text{pr}}{-}1$ values in the range $\{s{-}1,s,s{+}1\}$. In this first part of the synchronization procedure, the receiver performs a coarse time synchronization by discarding $N {-} \hat{s}_{\text{pr}}$ samples from its buffer. This way the buffer now contains $\hat{N}_{\text{pr}}{-}2$ symbols with a value of $\hat{s}=0$ or $\hat{s}=\pm 1$. This coarse synchronization is necessary as a first step in order to later apply the estimation algorithms for the fractional offsets. It is important to note that after this coarse synchronization, the receiver will still be misaligned in time, since the integer part of the CFO, i.e., $L_\text{CFO}$, also affects the value of $\hat{s}_{\text{pr}}$ and therefore results in a time misalignment. Moreover, the estimation of $\hat{s}_{\text{pr}}$ has been performed under the effect of the combination of the fractional parts $\lambda_\text{STO}{-}\lambda_{\text{CFO}}$, which can lead to an additional time offset of $\pm1$ samples.

\subsubsection{Estimation and Compensation of $L_{\text{STO}}$ and $L_{\text{CFO}}$}
A LoRa receiver can operate with the time misalignment resulting from $L_\text{CFO}$, since it translates into a frequency offset which finally compensates for $L_\text{CFO}$~\cite{Ghanaatian2019}. Ideally, however, the receiver should avoid the time misalignment due to $L_\text{CFO}$ because it results in inter-symbol interference~\cite{Ghanaatian2019}. Instead, $L_\text{CFO}$ should be compensated by introducing a frequency shift.
A simple and effective way of distinguishing the integer parts of the STO and the CFO has been described in~\cite{Bernier2019,Xhonneux2019}, where the receiver takes advantage of the $2.25$ downchirps in the preamble to separate the $L_\text{STO}$ and $L_{\text{CFO}}$ values. We implement the approach of~\cite{Bernier2019}, or equivalently~\cite[Eq. (26), (27)]{Xhonneux2019} in our LoRa SDR receiver for this part of the synchronization process. We re-synchronize our receiver in time, using now only the value of $L_\text{STO}$, and we compensate the effect of $L_{\text{CFO}}$ by introducing a frequency shift through multiplication with a complex exponential signal $e^{{-}j2\pi n \frac{L_{\text{CFO}}}{f_{s}}}$.

\subsubsection{Estimation and Compensation of $\lambda_{\text{STO}}$ and $\lambda_{\text{CFO}}$} \label{sec:lambda_est}

The estimation of a fractional offset in the frequency domain is a well-studied problem~\cite{Yang2011}. In particular, interpolation between the three maximum peaks of a sinc kernel can be used in order to find the value of the fractional offset with good accuracy and low-complexity. We propose two variations of the rational combination of the  three spectral lines (RCTSL) method described in~\cite{Yang2011}, to estimate $\lambda_{\text{CFO}}$ and $\lambda_{\text{STO}}$. 

As explained in~\cite{Xhonneux2019}, $\lambda_{\text{CFO}}$ has to be estimated and compensated before estimating $\lambda_{\text{STO}}$. For the estimation of $\lambda_{\text{CFO}}$ we use the $N_\text{pr}-2$ preamble symbols which were left in the buffer after the previous synchronization steps. %To err on the side of caution, we avoid using the last preamble symbol.
We dechirp the symbols and we perform a DFT of length $2(N_\text{pr}-2)N$ (the upsampling by a factor of two is required by the method of~\cite{Yang2011}) on an entire block of the $N_\text{pr}-2$ preamble symbols left in the buffer, i.e., $\tilde{\mathbf{Y}} = \text{DFT}\left(\tilde{\mathbf{y}} \odot \tilde{\mathbf{x}}_{0}^{*}\right)$, where $\tilde{\mathbf{y}} = [ \mathbf{y_{1}} \; \hdots \; \mathbf{y_{N_\text{pr}-2}} \; \mathbf{0}_{(N_\text{pr}-2)N}]$, $\tilde{\mathbf{x}}_{0} = [\mathbf{x}_{0} \; \hdots \; \mathbf{x}_{0} \; \mathbf{0}_{(N_\text{pr}-2)N}]$, and $\mathbf{0}_{(N_\text{pr}-2)N}$ denotes a zero vector of length $(N_\text{pr}-2)N$. Let $\tilde{Y}_{k_{\text{max}}}$ be the value of the maximum bin of the DFT $\tilde{\mathbf{Y}}$. Then, we compute~\cite{Yang2011}
\begin{align}
k_{\alpha} &= \frac{N}{\pi}\frac{|\tilde{Y}_{k_{\text{max}}{+}1}|^2 - |\tilde{Y}_{k_{\text{max}}{-}1}|^2}{u\left(|\tilde{Y}_{k_{\text{max}}{+}1}|^2 - |\tilde{Y}_{k_{\text{max}}{-}1}|^2\right) + v|\tilde{Y}_{k_{\text{max}}}|^2}, \label{eq:ka}
\end{align}
where $u = \frac{64N}{\pi^5+32\pi}$ and $v = u\frac{\pi^2}{4}$~\cite{Yang2011}. The fractional carrier frequency offset is computed as
\begin{align}
\lambda_{\text{CFO}} &= \frac{k_{\text{max}} + k_{\alpha}}{2(N_{\text{pr}}-2)} \mod 1.
\end{align}
Finally, the offset $\lambda_{\text{CFO}}$ is corrected through a multiplication of the received signal with $e^{{-}j2\pi n \frac{\lambda_{\text{CFO}}}{f_{s}}}$.

After $\lambda_{\text{CFO}}$ has been corrected, we re-use the $N_\text{pr}-2$ preamble symbols that were used for the estimation of $\lambda_{\text{CFO}}$ to estimate $\lambda_{\text{STO}}$ as follows. For every dechirped symbol we perform a length-$2N$ DFT (again, the upsampling by a factor of two is required by the method of~\cite{Yang2011}) and we combine the DFTs of the individual symbols in the following way
\begin{align}
|Y'_{k}|^2 = \sum_{i=1}^{N_\text{pr}-2} |Y^{(i)}_{k}|^2, \quad k \in \{0,\hdots,2N-1\}.
\end{align}
We then find $|Y'_{k_{\max}}|^2$ and we use it instead of $|\tilde{Y}_{k_{\max}}|^2$ in~\eqref{eq:ka} to calculate $k_{\alpha}$. The fractional sampling time offset is
\begin{align}
\lambda_{\text{STO}} &= \frac{k_{\text{max}} + k_{\alpha}}{2} \mod 1.
\end{align}
Finally, $\lambda_{\text{STO}}$ is compensated using time-domain interpolation.

\section{LoRa Testbed and Measurement Results}
In this section we will briefly describe our GNU Radio LoRa PHY implementation. Moreover, we will use this implementation in a testbed in order to experimentally assess the error rate performance of LoRa. Our open-source implementation is publicly available for the research and development of algorithms that improve the performance of LoRa receivers~\cite{TCLoRaRepo}.

\subsection{GNU Radio LoRa PHY Implementation}
In the GNU Radio implementation of the LoRa Tx and Rx chains the user can choose all the parameters of the transmission, such as the spreading factor, the coding rate, the bandwidth, the presence of a header and a CRC, the message to be transmitted, etc. In the Tx chain, the implementation contains all the main blocks of the LoRa transceiver described in Section~\ref{sec:loraphy}, i.e., the header- and the CRC-insertion blocks, the whitening block, the Hamming encoder block, the interleaver block, the Gray mapping block, and the modulation block. On the receiver side there is the packet synchronization block, which performs all the necessary tasks needed for the synchronization, such as the STO and CFO estimation and correction as described in Section~\ref{sec:frameSynch}. The demodulation block follows, along with the Gray demapping block, the deinterleaving block, the Hamming decoder block and the dewhitening block, as well as a CRC block.

\subsection{Testbed description}

\begin{figure}[t]
	\centering
	\includegraphics[width=0.45\textwidth]{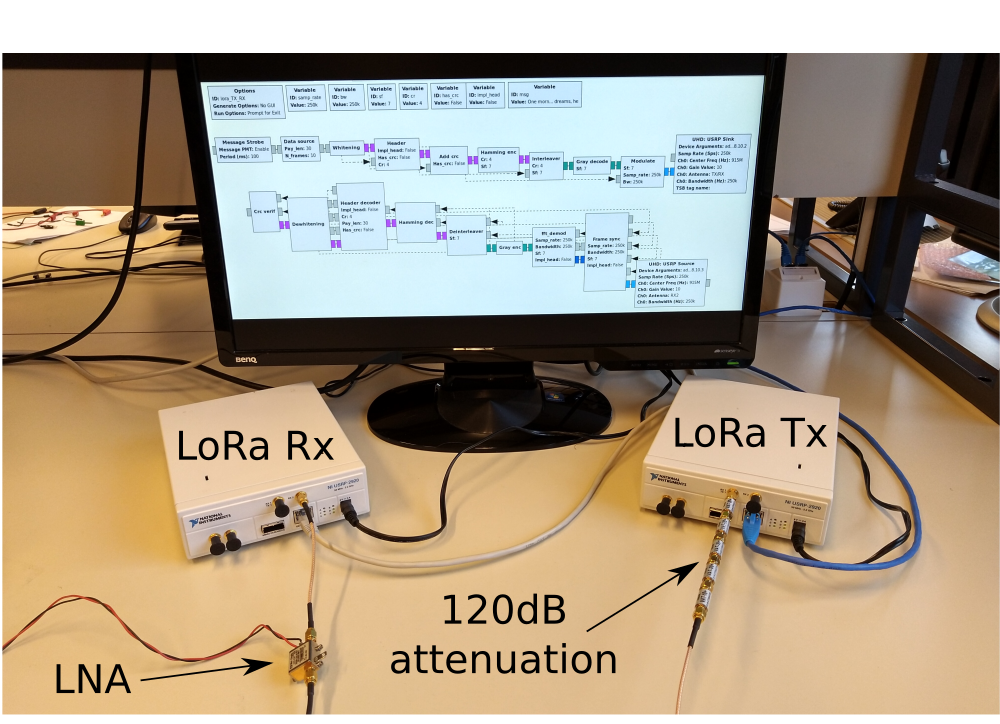}
	\caption{LoRa testbed in USRPs using our GNU Radio SDR platform.}
	\label{fig:LoRaTestbedPhoto}
	\vspace{-0.25cm}
\end{figure}
Our testbed uses National Instruments (NI) 2920 USRP transceivers~\cite{NIUSRPdatasheet}, but any SDR device that supports GNU Radio can be used instead. The transmit power of the NI 2920 USRP ranges from $-11$~dBm to $20$~dBm. The nominal carrier frequency used for our transmissions is $915$~MHz. We note that, as can be seen in Fig.~\ref{fig:LoRaTestbedPhoto}, in order to avoid interference from other sources in 915 MHz ISM band, we use a cable with attenuators for transmission instead of antennas.  The carrier frequency is generated within each USRP from a reference clock. This reference clock can be shared between the Tx and the Rx USRPs, leading to $L_{\text{CFO}} = \lambda_{\text{CFO}} = 0$, for initial validation experiments without CFO and to obtain the CFO ground truth for measurements of the CFO compensation.\footnote{We note that, even when sharing the reference clock, a random, time-varying STO is present, since only the frequencies, not the phase, are locked.} We use the Adafruit Feather 32u4 RFM95 as a commercial LoRa transceiver~\cite{AdaFruit} and an external MiniCircuits ZX60-33LN+ low-noise amplifier~\cite{LNA} with a gain of approximately $19$~dB at the receiver of the USRP.  Using this testebd, we have successfully transmitted LoRa packets from a USRP to a USRP, from a commercial LoRa transceiver to a USRP, and from a USRP to a commercial LoRa transceiver.

\subsection{Testbed results}
\begin{figure}
  \centering
  \begin{tikzpicture}

	%\pgfplotsset{grid style={dashed}}
	\small

	\begin{semilogyaxis}[
		width = \figurewidth\columnwidth,
		height = \figureheight\columnwidth,
		xlabel = {SNR (dB)},
		ylabel = {Bit Error Rate},
		label style={font=\small},
    tick label style={font=\footnotesize},
		ylabel near ticks,
		xlabel near ticks,
		xmin = -13, xmax = -6,
		ymin = 1e-5, ymax = 1,
		grid = both,
		legend image post style={scale=0.6},
		%legend style={at={(0.45,-0.25)},anchor=north,font=\tiny},
		%legend style={font=\tiny},
		%legend cell align={left},
		%legend columns={3},
		%transpose legend,
	]

		%%% SF = 7 %%%

		%%%%% Uncoded %%%%%%%
		%% Monte Carlo simulation %%
		% lambda = 0.0
		\addplot[black, ultra thick, solid, mark=none, mark options={scale=1}] table[x index=0, y index = 1] {figs/data/AWGN_uncoded.dat};
		\label{lambda0.0MC}

		%% Testbed %%
		% without CFO
		\addplot[Set1-7-2, thick, solid, mark=*, mark options={scale=1}] table[x index=0, y index = 1] {figs/data/BER_cfo0_cr0.dat};
		\label{lambda0.0TB}
		% with CFO
		\addplot[Set1-7-5, thick, solid, mark=square*, mark options={scale=1, solid}] table[x index=0, y index = 1] {figs/data/BER_cfo03_cr0.dat};
		\label{lambda0.3TB}

		%%%%% Coded %%%%%%%		
		
		%% Monte Carlo simulation %%
		% lambda = 0.0
		\addplot[black, ultra thick, densely dashed, mark=none, mark options={scale=1}] table[x index=0, y index = 1] {figs/data/BER_RES_SF7_F72_lambda0.00_CWLen8_Interl1_Gray1_Reverse1_Reduced0_0.dat};
		\label{Coded_lambda0.0MC}

		%% Testbed %%
		% without CFO
		\addplot[Set1-7-2, thick, densely dashed, mark=*, mark options={scale=1, solid}] table[x index=0, y index = 1] {figs/data/BER_cfo0_cr4.dat};
		\label{Coded_lambda0.0TB}
		% with CFO
		\addplot[Set1-7-5, thick, densely dashed, mark=square*, mark options={scale=1, solid}] table[x index=0, y index = 1] {figs/data/BER_cfo03_cr4.dat};
		\label{Coded_lambda0.3TB}

		% Draw first "Legend" node using a left justified shortstack, position using relative axis coordinates
		\node [draw,fill=white,inner sep=2pt] at (rel axis cs: 0.77,0.86) {
		\tiny
		\setlength{\tabcolsep}{2pt}
		\begin{tabular}{lc}
			\multicolumn{2}{c}{\scriptsize\underline{Uncoded LoRa (SF$=7$)}} \\		
			Simulation (AWGN):	& \ref{lambda0.0MC} \\
			Testbed (STO): & \ref{lambda0.0TB} \\
			Testbed (STO \& CFO): & \ref{lambda0.3TB}
		\end{tabular}};
		
		% Draw second "Legend" node using a left justified shortstack, position using relative axis coordinates
		\node [draw,fill=white,inner sep=2pt] at (rel axis cs: 0.28,0.135) {
		\tiny
		\setlength{\tabcolsep}{2pt}
		\begin{tabular}{lc}
			\multicolumn{2}{c}{\scriptsize\underline{Coded LoRa (SF$=7$, \text{CR} = \nicefrac{4}{8})}} \\		
			Simulation (AWGN):	& \ref{Coded_lambda0.0MC} \\
			Testbed (STO): & \ref{Coded_lambda0.0TB} \\
			Testbed (STO \& CFO): & \ref{Coded_lambda0.3TB}
		\end{tabular}};

	\end{semilogyaxis}

\end{tikzpicture}%
  \caption{Uncoded and coded LoRa BER for $\text{SF} = 7$ (testbed vs simulation).}
  \label{fig:BER_uncoded_sf7}
  \vspace{-0.25cm}
\end{figure}
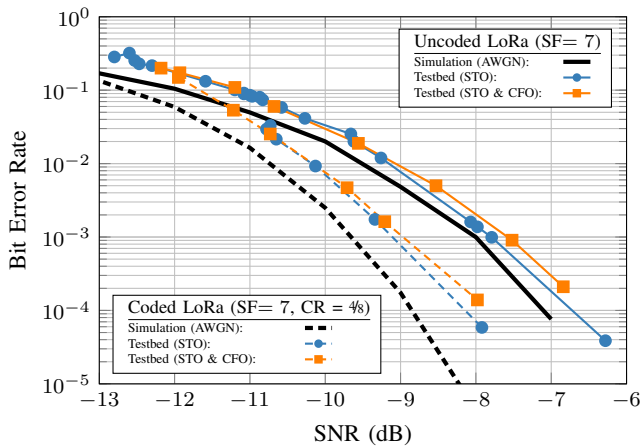

In Fig.~\ref{fig:BER_uncoded_sf7} we show bit error rate (BER) results for $\text{SF}=7$ and for a payload of $64$ bytes obtained from transmissions using our USRP testbed which runs our GNU Radio implementation of LoRa PHY for both the uncoded and coded ($\text{CR} = \nicefrac{4}{8}$) cases. The experimental curves are compared to the performance of LoRa obtained through MATLAB-based Monte Carlo simulations. Results for the cases with and without CFO are shown. Low SNR values are achieved by using multiple attenuators, reaching a total attenuation of $120$~dB. We generate different SNR values by fixing the attenuation to $120$~dB and sweeping the Tx gain. Even for a fixed Tx gain, we observed that the received signal power may vary between different runs. Therefore, for a given SNR we consider only the packets with received signal power no more than $\pm 1$~dB of the mean received signal strength value for the given Tx gain. The SNR for the experimental curves is measured as the ratio between the power of the maximum DFT bin over the power contained in the rest of the bins after synchronization and STO/CFO compensation. 

For uncoded LoRa, we observe that in the case with STO, but without CFO, the experimental curve matches the simulated performance of uncoded LoRa under AWGN~very well~\cite{Elshabrawy2018,Afisiadis2019b}. This result shows that the STO estimation and compensation algorithm described in Section~\ref{sec:frameSynch} is accurate, leading to a testbed performance that is very close to the theoretical limits for AWGN. Moreover, we observe that in the case where the local oscillator is not shared, and therefore both STO and CFO are present, the performance degradation of the implementation compared to the bound given by the AWGN simulation curve is still less than $1$dB.

For coded LoRa, we observe that in the case with STO, but without CFO, the experimental curve is within $1$~dB of the simulated performance of coded LoRa under AWGN~\cite{Afisiadis2019c}. Additionally, we observe that the performance degradation in the presence of both STO and CFO is almost negligible, since the CFO estimation and compensation works well, and moreover, coded LoRa is relatively resistant to small residual CFO values after the correction~\cite{Afisiadis2019c}.

\section{Conclusion}
In this work we present our open-source GNU Radio implementation of LoRa PHY, along with the description of the employed synchronization and offset correction algorithms. Validation measurements show that the testbed error rate performance is within $1$~dB of MATLAB simulations, even with fully decoupled transmitters and receivers. Our SDR implementation can be used in testbeds that give very stable results, and can thus be a useful tool on the experimental validation of proposed algorithms for improved LoRa transceivers.

\section*{Acknowledgment}
The authors thank Pieter Robyns and Mathieu Xhonneux for helpful discussions on the GNU Radio LoRa implementation of~\cite{Robyns2018} and on LoRa packet synchronization, respectively, as well as Pascal Giard for helpful discussions on LoRa PHY reverse engineering and GNU Radio implementation. 


\bibliographystyle{IEEEtran}
\bibliography{refs}

\end{document}